\documentclass[twocolumn,aps]{revtex4}

\def\Sbar{S\llap{$\overline{\phantom{I}}$}}
\def\half{{\textstyle{1\over2}}}
\def\eq#1{eq.$\,$(\ref{#1})}

\def\Eq#1{Eq.$\,$(\ref{#1})}

\begin{document}

\title{Subjective and Objective Probabilities in Quantum Mechanics}
\author{Mark Srednicki}
\email{mark@physics.ucsb.edu}
\affiliation{
Department of Physics,
University of California,
Santa Barbara, CA 93106 USA 
}

\begin{abstract}
We discuss how the apparently objective probabilities predicted by quantum mechanics
can be treated in the framework of Bayesian probability theory, in which all probabilities
are subjective.  Our results are in accord with earlier work by Caves, Fuchs, and Schack,
but our approach and emphasis are different.  We also discuss the problem of choosing a noninformative prior for a density matrix. 
\end{abstract}

\maketitle

\section{Introduction}
\label{intro}

Probability plays a central role throughout human affairs,
and so everyone has an intuitive idea of what it is.
Moreover, because of the extreme generality and widespread use
of the concept of probability, 
it cannot be easily defined in terms of anything more basic.
For example, the dictionary that I have in my office,
{\it Webster's Ninth New Collegiate},
says that {\it probability\/} is
``the state or quality of being probable'';
that to be {\it probable\/} is to be
``supported by evidence strong enough to establish presumption
but not proof''; and that {\it presumption\/} is 
``the ground, reason, or evidence lending probability to a belief''.
This is clearly unhelpful to anyone who does not already know
what probability is.

In mathematics and physics, we are often faced with a
concept that is both simple enough to be clearly understood,
and fundamental enough to resist definition; for example,
a {\it straight line\/} in euclidean geometry.
To make progress, we do not attempt to devise
ever clearer definitions, but instead formulate axioms that
our understood but undefined objects are postulated to obey. 
Then, using codified rules of logical inference, we prove
theorems that follow from the axioms.   

It is instructive to treat probability as one of these primitive concepts.
Dispensing, then, with any attempt at definition, we say that
the {\it probability\/} that a {\it statement\/} is {\it true\/} is
a real number between zero and one.  A statement may be true or false;
if we know it to be true, we assign it a probability of one, and
if we know it to be false, we assign it a probability of zero.
If we do not know whether it is true or false, we assign it 
a probability between zero and one.  

There is typically no definitive way 
to make this assignment.  Different people could (and often do) assign 
different numerical values to the probability that some particular
statement (``the stock price of Microsoft will be higher one year from now'')
is true.  In this sense, probability is {\it subjective}.
This point of view is {\it Bayesian}.

Probability also enters quantum mechanics, in a seemingly more fundamental way.  For example, given a wave function $\psi(x,t)$ for a particle in one
dimension, the rules of quantum mechanics 
(which are apparently laws of nature) 
tell us that we must assign a probability $|\psi(x,t)|^2 \,dx$ to the statement 
``at time $t$, the particle is between $x$ and $x+dx$''.
Different people do not appear to have a choice about this assignment.
In this sense, quantum probability appears to be {\it objective}.  
 
The goal of this paper is to understand the how the apparently objective
probababilities of quantum mechanics can be fit into
the Bayesian framework, which allows different people
to make different probability assignments.
This issue has been addressed before by Caves, Fuchs, and Schack \cite{cfs},
and our results are in broad agreement with theirs.  
However, we emphasize a somewhat different approach to certain
issues that we will explain as we go along.

In section \ref{axioms},  in order to fix the notation and key concepts,
we briefly review the axioms and basic theorems of probability theory.
In section \ref{probprob}, we introduce the notion of a probability of a
probability, and explain how it can be applied to experimental data
to turn an originally subjective probability into an increasingly objective
one, in the sense that all but strongly biased observers agree with
the final probability assignment.  In section \ref{qm}, we apply this
formalism to the probabilities of quantum mechanics.  In section \ref{pdmdm},
we discuss when and why it is preferable to assign probabilities
to possible density matrices for a quantum system, rather than assigning
a particular density matrix.
In section \ref{nonin}, we discuss the construction of 
noninformative prior distributions for density matrices.  
We summarize and conclude in section \ref{con}.

\section{The Axioms of Probability}
\label{axioms}

The statements 
to which we may assign probabilities must obey a logical calculus.
Some key definitions (in which ``iff'' is short for ``if and only if''):

$S =$ a statement.

$\Omega =$ a statement known to be true.

$\emptyset= $ a statement known to be false.

$\Sbar =$ a statement that is true iff $S$ is false.

$S_1 \vee S_2 =$ a statement that is true iff
either $S_1$ or $S_2$ is true.

$S_1 \wedge S_2 =$ a statement that is true iff
both $S_1$ and $S_2$ are true.

$S_1$ and $S_2$ are {\it mutually exclusive\/} iff
                             $S_1 \wedge S_2 = \emptyset.$ 

$S_1, \ldots, S_n$ are a {\it complete set\/} iff
            $S_1 \vee \ldots \vee S_n = \Omega$ and
            $S_i \wedge S_j = \emptyset$ for $i\ne j$. 

Elementary logical relationships among statements include
$S \vee \Sbar = \Omega$,
$S \wedge \Sbar = \emptyset$,
$S \wedge \Omega = S$, 
$S_1\wedge(S_2\vee S_3) =  (S_1\wedge S_2)\vee(S_1\wedge S_3)$, 
etc.
Denoting the probability assigned to a statement $S$ as $P(S)$, 
we can state the first three axioms of probability.

Axiom 1. $P(S)$ is a nonnegative real number.

Axiom 2. $P(S) = 1$ iff $S$ is known to be true.
Axiom 3. If $S_1$ and $S_2$ are mutually exclusive, 
then $P(S_1 \vee S_2) = P(S_1) + P(S_2)$. 

From these axioms, and the logical calculus of statements,
we can derive some simple lemmas:

Lemma 1. $P(\Sbar\,) = 1 - P(S).$ 

Lemma 2. $P(S)  \le 1.$ 

Lemma 3. $P(S)=0$ iff $S$ is known to be false.

Lemma 4. $P(S_1 \wedge S_2) = P(S_1) + P(S_2) - P(S_1 \vee S_2).$ 

\noindent
We omit the proofs, which are straightforward.

We will also need the notion of a {\it conditional\/} statement
$S_2|S_1$.   $S_2|S_1$ is a statement 
if and only if $S_1$ is true; otherwise, $S_2|S_1$ is {\it not\/} 
a statement, and cannot be assigned a probability.  
Given that $S_1$ is true,
the statement $S_2|S_1$ is true if and only if $S_2$ is true.
The probability that $S_2|S_1$ is true is then specified by

Axiom 4.  $P(S_2|S_1) =  P(S_1 \wedge S_2)/P(S_1).$

\noindent
Note that, if $P(S_1)=0$, then $S_1=\emptyset$ by Lemma~3, 
and so both sides of Axiom~4 are undefined: 
the right side because we have divided by zero, and 
the left side because $S_2|\emptyset$ is not a statement.

Another concept we will need is that of {\it independence\/} between statements.
Two statements are said to be {\it independent\/} 
if the knowledge that one of them if true
tells us nothing about whether or not the other one is true. 
Thus, if $S_1$ and $S_2$ are independent, we should have
$P(S_1|S_2)=P(S_1)$ and $P(S_2|S_1)=P(S_2)$.  Using these relations
and Axiom~4, we get a result that can be used
as the definition of independence,

$S_1$ and $S_2$ are {\it independent\/} if and only if 
$P(S_1 \wedge S_2) = P(S_1)  P(S_2)$.

Note that independence is a property of probability assignments, rather than the
statements themselves.  Thus, people can disagree on whether or not two
statements are independent.

\section{Probabilities of probabilities}
\label{probprob}

What limitations, 
if any, should be placed on the nature of statements to which
we are allowed to assign probabilities?

There are various schools of thought.
{\it Frequentists\/} assign probabilities only
to {\it random variables}, a highly restricted class of statements that we shall
not attempt to elucidate.  
{\it Bayesians} allow a wide range of statements,
including statements about the future
such as ``when this coin is flipped it will come up heads,''
statements about the past such as ``it rained here yesterday,'' and
timeless statements such as ``the value of Newton's constant is between
$6.6$ and $6.7 \times 10^{-11}\,$m$^3$/kg$\,$s$^2$.''   Some level of
precision is typically insisted on, so that, for example, ``red is good'' 
might be rejected as too vague.

A major thesis of this paper is that the class of allowed statements 
should include statements about the probabilities of other statements.  
Some Bayesians (for example, de Finetti \cite{definpp}) 
reject this concept as meaningless.   However,
it has found some acceptance and utility in decision theory, where it is sometimes called a {\it second order
probability\/}; see, e.g., \cite{dec}.  In particular, it is an experimental
fact that people's decisions depend not only on the probabilities they assign to various
alternatives, but also on the degree of confidence that they have in their own
probability assignments \cite{dec}.   This degree of confidence can be quantified
and treated as a probability of a probability.  

To illustrate how we will use the concept, consider the following problem.
Suppose that we have a situation with exactly two possible outcomes 
(for example, a coin flip).  Call the two outcomes 
$A$ and $B$.   In the terminology of the logical calculus, $A\vee B=\Omega$
and $A\wedge B=\emptyset$, so that $A$ and $B$ are a complete set.
The probability axioms then require $P(A)+P(B)=1$, but do not tell us
anything about either $P(A)$ or $P(B)$ alone.

In the absence of any other information, we invoke Laplace's 
{\it principle of insufficient reason} (also called the {\it principle of
indifference}):
when we have no cause to prefer one statement over another, 
we assign them equal probabilities.
Thus we are instructed to choose $P(A)=P(B)=\half$.
While this assignment is logically sound,  
we clearly cannot have a great deal
of confidence in it; typically, we are prepared to
abandon it as soon as we get some more information. 

Another (and, we argue, better) strategy is to retreat 
from the responsibility of assigning a particular value to $P(A)$, 
and instead assign a probability $P(H)$ to the statement
$H = {}$``the value of $P(A)$ is between $h$ and $h+dh$.''  
Here $dh$ is infinitesimal, and $0\le h\le 1$.  
Then $P(H)$ takes the form $p(h)dh$, where $p(h)$ is a nonnegative function
that we must choose, normalized by $\int_0^1 p(h)dh = 1$.  
We might choose $p(h)=1$, for example.  

Now suppose we get some more information about $A$ and $B$.  
Suppose that the situation that produces either $A$ or $B$ as an 
outcome can be recreated repeatedly (each repetition will be called a {\it trial\/}),
and that the outcomes of the different trials are (we believe) independent.
Suppose that the result of the first $N$ trials is $N_{\!A}$ $A$'s and 
$N_{\!B}$ $B$'s, in a particular order. What can we say now?

The formula we need is

Bayes' Theorem.  $P(H|D) =P(D|H)P(H)/P(D).$

\noindent
Bayes' theorem follows immediately from Axiom~4; since $H\wedge D$ 
is the same as $D\wedge H$, we have
$P(H|D)P(D) = P(H\wedge D) = P(D|H)P(H)$.
While $H$ and $D$ can be any allowed statements,
the letters are intended to denote ``hypothesis'' and ``data''.
Bayes' theorem tells us that, given a hypothesis $H$ to which
we have somehow assigned a {\it prior probability\/} $P(H)$ 
(whether by the principle of indifference, or by any other means), 
and we know (or can compute)
the {\it likelihood\/} $P(D|H)$ of getting a particular set of data $D$ given
that the hypothesis $H$ is true, then we can compute
the {\it posterior probability\/} $P(H|D)$ that the hypothesis $H$ is true,
given the data $D$ that we have obtained.  
Furthermore, if we have a complete set of hypotheses $H_i$, then we can
express $P(D)$ in terms of the associated likelihoods and
prior probabilities: starting with
$D=D\wedge\Omega=D\wedge(H_1\vee H_2\vee\ldots)
=(D\wedge H_1)\vee(D\wedge H_2)\vee\ldots,$
and noting that $D\wedge H_i$ and $D\wedge H_j$ are mutually
exclusive when $i\ne j$, we have
\begin{equation}
P(D)=\sum_i P(D\wedge H_i)   = \sum_i P(D|H_i)P(H_i) ,
\label{pofd}
\end{equation}
where the first equality follows from Axiom~3, and the second from Axiom~4.

To apply these results to the case at hand, 
recall that our hypothesis is $H={}$``$P(A)$ is between $h$ and $h+dh$''.
We have assigned this hypothesis a prior probability $P(H)=p(h)dh$.
The data $D$ is a string of $N_{\!A}$ $A$'s and $N_{\!B}$ $B$'s,
in a particular order; each of the $N=N_{\!A}+N_{\!B}$ outcomes is 
assumed to be independent of all the others.
Using the definition of independence, we see that the likelihood is
\begin{eqnarray}
P(D|H) &=& P(A)^{N_{\!A}} P(B)^{N_{\!B}} 
\nonumber \\
\noalign{\medskip}
&=&  h^{N_{\!A}}(1-h)^{N_{\!B}}.
\label{pdh}
\end{eqnarray}
Applying Bayes' Theorem, we get the posterior probability
\begin{equation}
P(H|D)= P(D)^{-1} h^{N_{\!A}}(1-h)^{N_{\!B}}p(h)dh,
\label{bayesab}
\end{equation}   
where 
\begin{equation}
P(D)=\int_0^1 h^{N_{\!A}}(1-h)^{N_{\!B}}p(h)dh.
\label{bayespd}
\end{equation}   
If the number of trials $N$ is large, and if the prior probability $p(h)$ has been
chosen to be a  slowly varying function, then the posterior probability $P(H|D)$ 
has a sharp peak at $h=h_{\rm exp} \equiv N_{\!A}/N$, the fraction of trials that resulted in 
outcome $A$.  The width of this peak is proportional to
$N^{-1/2}$ if both $N_{\!A}$ and $N_{\!B}$ are large, and to
$N^{-1}$ if either $N_{\!A}$ or $N_{\!B}$ is small (or zero).
Thus, after a large number of trials, we can be confident
that the probability $P(A)$ that the next outcome will be $A$
is close to the fraction of trials that have already resulted in $A$.
The only people who will not be convinced of this are those whose
choice of prior probability $p(h)$ is strongly biased against the value $h=h_{\rm exp}$.
Thus, the value $h_{\rm exp}$ for the probability $h$ is becoming {\it objective}, 
in the sense that almost all observers agree on it.  
Furthermore, those who do not
agree can be identified {\it a priori\/} by noting that their prior probabilities are 
strong functions of $h$.

Those who reject the notion of a probability of a probability, but who
accept the practical utility of this analysis (which was originally carried
out by Laplace), have two options.  
Option one is to declare that $h$ is not actually a probability; it is
rather a {\it limiting frequency\/} or a {\it propensity\/} or a {\it chance}.
Option two is to declare that $p(h)dh$ is not actually a probability;  
it is a {\it measure\/} or a {\it generating function}.  

Let us explore option two in more detail.
Rather than assigning a second-order probability to 
$H={}$``$P(A)$ is between $h$ and $h+dh$'', we assign a probability
to {\it every finite sequence\/} of outcomes; that is, we choose values
for $P(A)$, $P(B)$, $P(AB)$, $P(BA)$, $P(AAA)$, $P(AAB)$,
and so on, for strings of arbitrarily many outcomes.  
We assume that all possible strings of $N$ outcomes form a complete set.
Our probability assignments must of course satisfy the probability axioms,
so that, for example, $P(A)+P(B)=1$.  We also insist that the assignments
be {\it symmetric}; that is, independent of the ordering of the outcomes,
so that, for example,
\begin{equation}
P(AAB)=P(ABA)=P(BAA).
\label{paab}
\end{equation}
Furthermore, the assignments for strings of $N$ outcomes must be consistent with those for $N+1$ outcomes; this means that, 
for any particular string of $N$ outcomes $S$,
\begin{equation}
P(S)=P(SA)+P(SB).
\label{psab}
\end{equation}
A set of probability assignments that satisfies these requirements 
is said to be {\it exchangeable}.  Then, the 
{\it de Finetti representation theorem\/} \cite{defin}
states that, given an exchangeable set
of probability assignments for all possible strings of outcomes, 
the probability of getting a specific string $D$ of $N$ outcomes 
that includes exactly
$N_{\!A}$ $A$'s and $N_{\!B}$ $B$'s can always be written in the form
\begin{equation}
P(D)=\int_0^1 h^{N_{\!A}}(1-h)^{N_{\!B}}p(h)dh,
\label{defin}
\end{equation}
where $p(h)$ is a unique nonnegative function that obeys the 
normalization condition $\int_0^1 dh\,p(h)=1$, and is the same
for every string $D$.  
Note that \eq{defin} is exactly the same as \eq{bayespd}.
Thus an exchangeable probability assignment to sequences
of outcomes can be characterized by a function $p(h)$ that
can be (as we have seen) consistently treated as a probability of a probability.
But those who find this notion unpalatable are free to think of
$p(h)$ as specifying a measure, or a generating function, or a similar euphemism.

To summarize, if we need to assign a prior probability but have little information, it can be more constructive to abjure, and instead assign a probability to a range
of possible values of the needed prior probability.  
This probability of a probability can then be updated with 
Bayes' theorem as more information comes in.  
 
\section{Probability in Quantum Mechanics}
\label{qm}

Suppose we are given a qubit: a quantum system with a two-dimension Hilbert
space.  (We will use the language appropriate to a spin-one-half particle to describe it.)
We are asked to make a guess for its quantum state.

Without further information, the best we can do is invoke 
the principle of indifference.  In the case of a finite set of possible
outcomes, this principle is based on the permutation symmetry of the outcomes;
we choose the unique probability assignment that is invariant under this symmetry.
The quantum analog of the permutation of outcomes is the unitary symmetry of 
rotations in Hilbert space. The only quantum state that is invariant under this symmetry is
the fully mixed density matrix
\begin{equation}
\rho = \half I.
\label{rhomix}
\end{equation}
Thus we are instructed to choose \eq{rhomix} as the quantum state of the system.
While this assignment is logically sound,  
we clearly cannot have a great deal
of confidence in it; typically, we are prepared to
abandon it as soon as we get some more information. 

Another (and, we argue, better) strategy is to retreat 
from the responsibility of assigning a particular state (pure or mixed)
to the system, and instead assign a probability $P(H)$ to the statement
$H = {}$``the quantum state of the system is a density matrix within a volume 
$d\rho$ centered on $\rho$'', where $\rho$ is a particular $2\times 2$ hermitian 
matrix with nonnegative eigenvalues that sum to one,
and $d\rho$ is a suitable differential volume element in the space of such matrices.
We can parameterize $\rho$ with three real numbers $x$, $y$, and $z$ via
\begin{equation}
\rho ={1\over2} \pmatrix{ 1+z & x- iy \cr
\noalign{\smallskip}
 x+iy & 1-z \cr},
\label{rhoxyz}
\end{equation}
where where $x^2 + y^2 + z^2 \equiv r^2 \le 1$.  We then take
$d\rho =dV$, where $dV=(3/4\pi)dx\,dy\,dz$ is the normalized volume element:
$\int dV = 1$.
$P(H)$ takes the form $p(\rho)dV$, where $p(\rho)$ is a nonnegative function that we must choose, normalized by $\int p(\rho)dV = 1$.  
We might choose $p(\rho)=1$, for example.   

Now suppose we get some more information about the quantum state of the 
system.  Suppose that the procedure that prepares the quantum state of the 
particle can be recreated repeatedly
(each repetition of this will be called a {\it trial\/}),
and that the outcomes of measurements performed on each prepared
system are (we believe) independent.  Suppose further
that we have access to a Stern--Gerlach apparatus that allows us to measure
whether the spin is $+$ or $-$ along an axis of our choice.  
We choose the $z$ axis.  Suppose that the result of the first $N$ trials is 
$N_{+}$ $+$'s and $N_{-}$ $-$'s.
What can we say now?

Given a density matrix $\rho$, parameterized by \eq{rhoxyz}, 
the rules of quantum mechanics tell us that the probability
that a measurement of the spin along the $z$ axis will yield $+1$ is 
\begin{equation}
P(\sigma_z\,{=}\,{+}1|\rho) = {\rm Tr}\,\half(1+\sigma_z)\rho = \half(1+z) ,
\label{PSzp}
\end{equation}
where $\sigma_z$ is a Pauli matrix,
and the probability that this measurement will yield $-1$ is 
\begin{equation}
P(\sigma_z\,{=}\,{-}1|\rho) = {\rm Tr}\,\half(1-\sigma_z)\rho = \half(1-z) .
\label{PSzm}
\end{equation}

Now we use Bayes' theorem.  Our hypothesis is $H={}$``the quantum state
is within a volume $d\rho$ centered on $\rho$''.
We have assigned this hypothesis a prior probability $P(H)=p(\rho)d\rho$.
The data $D$ is a string of $N_+$ $+$'s and $N_-$ $-$'s,
in a particular order; each of the $N=N_++N_-$ outcomes is 
assumed to be independent of all the others.
Using the definition of independence, we see that the likelihood is
\begin{eqnarray}
P(D|H) &=& [P(\sigma_z\,{=}\,{+}1|\rho)]^{N_+}  [P(\sigma_z\,{=}\,{+}1|\rho)]^{N_-} 
\nonumber \\
\noalign{\medskip}
&=&  [\half(1+z)]^{N_+}  [\half(1-z)]^{N_-} .
\label{pdhqm}
\end{eqnarray}
Applying Bayes' Theorem, we get the posterior probability
\begin{equation}
P(H|D)= P(D)^{-1}  [\half(1+z)]^{N_+}  [\half(1-z)]^{N_-}p(\rho)d\rho,
\label{bayesabqm}
\end{equation}   
where 
\begin{equation}
P(D)=\int   [\half(1+z)]^{N_+}  [\half(1-z)]^{N_-}p(\rho)d\rho.
\label{bayespdqm}
\end{equation}   
When the number of trials $N$ is large, and the prior probability $p(\rho)$ is a 
slowly varying function, the posterior probability $P(H|D)$ has a sharp peak at
$z=z_{\rm exp}\equiv(N_+ - N_-)/N$. 
Thus, after a large number of trials in which we measure $\sigma_z$, 
we can be confident of the value of the parameter $z$ in the density
matrix of the system.
The only people who will not be convinced of this are those whose
choice of prior probability $p(\rho)$ is strongly biased against the value 
$z=z_{\rm exp}$.  Furthermore, those who do not
agree can be identified {\it a priori\/} by noting that their prior probabilities are 
strong functions of $\rho$.

We can of course orient our Stern--Gerlach apparatus along different
axes.  If we choose the $x$ axis or the $y$ axis, 
the relevant predictions of quantum mechanics are
\begin{eqnarray}
P(\sigma_x\,{=}\,{+}1|\rho) &=& {\rm Tr}\,\half(1+\sigma_x)\rho =  \half(1+x) ,
\label{PSxp} \\
\noalign{\medskip}
P(\sigma_x\,{=}\,{-}1|\rho) &=& {\rm Tr}\,\half(1-\sigma_x)\rho =  \half(1-x) ,
\label{PSxm} \\
\noalign{\medskip}
P(\sigma_y\,{=}\,{+}1|\rho) &=& {\rm Tr}\,\half(1+\sigma_y)\rho =  \half(1+y) ,
\label{PSyp} \\
\noalign{\medskip}
P(\sigma_y\,{=}\,{-}1|\rho) &=& {\rm Tr}\,\half(1-\sigma_y)\rho =  \half(1-y) .
\label{PSym} 
\end{eqnarray}
For each trial, we can choose whether to measure $\sigma_x$, $\sigma_y$, or $\sigma_z$. (We could also choose to measure along any other axis.)
Then, if the outcomes include $N_{+z}$ measurements of $\sigma_z$
with the result $\sigma_z=+1$, and so on,  the posterior probability becomes
\begin{eqnarray}
P(H|D)&=& P(D)^{-1}  [\half(1+x)]^{N_{+x}}  [\half(1-x)]^{N_{-x}}
\nonumber \\
&& {} \times [\half(1+y)]^{N_{+y}}  [\half(1-y)]^{N_{-y}}
\nonumber \\
&& {} \times [\half(1+z)]^{N_{+z}}  [\half(1-z)]^{N_{-z}}
p(\rho)d\rho, \qquad
\label{bayesabcqm}
\end{eqnarray}   
where $P(D)$ is given by the obvious integral.
Clearly the discussion in the preceding paragraph is simply
triplicated, and, when the number of trials is large,
we have determined the entire density matrix
to the satisfaction of all but strongly biased observers.
Our subjective probabilities of probabilities have led
us to an objective conclusion about quantum probabilities.

In \cite{cfs}, Caves et al arrived at an essentially identical result. 
The main difference in their analysis is that they regarded $p(\rho)d\rho$ 
as a measure rather than a probability.  
This approach required them to prove, first, a quantum version
of the de Finetti theorem \cite{qdef}, and, second, that Bayes' theorem can be
applied to $p(\rho)d\rho$ \cite{qbayes}.  Both steps become unnecessary if
we treat $p(\rho)d\rho$ as, fundamentally, a probability.

\section{Probabilities for Density Matrices vs.~Density Matrices}
\label{pdmdm}

If we assign an impure density matrix $\rho$ to a quantum system, does
this not already take into account our ignorance about it?  Why is
it preferable to assign, instead, a probability $p(\rho)d\rho$
to the set of possible density matrices?

It depends on the nature of our ignorance.  Suppose, for example,
the system is the spin of an electron plucked from the air.   Then we
expect that \eq{rhomix} will describe it, in the sense that if we do
repeated trials (plucking a new electron each time, and measuring
its spin along an axis of our choice), we will find that 
$x_{\rm exp}\equiv (N_{+x}-N_{-x})/(N_{+x}+N_{-x})$,
$y_{\rm exp}\equiv (N_{+y}-N_{-y})/(N_{+y}+N_{-y})$, and
$z_{\rm exp}\equiv (N_{+z}-N_{-z})/(N_{+z}+N_{-z})$ all tend to zero.

Suppose instead that the spin is prepared by a technician who (with
the aid of a Stern--Gerlach device) puts it in either a pure state with 
$\sigma_z=+1$, or a pure state with $\sigma_x=+1$, and each time
decides which choice to make by flipping a coin that we believe is fair.
In this case the appropriate density matrix is
\begin{eqnarray}
\rho &=& \half[\half(1+\sigma_z)] + \half[\half(1+\sigma_x)]
\nonumber \\
\noalign{\medskip}
&=& {1\over 4}\pmatrix{ 3 & 1 \cr 1 & 1 \cr}.
\label{rhoodd}
\end{eqnarray}
Comparing with \eq{rhoxyz}, we see that we now we expect 
$x_{\rm exp}$,
$y_{\rm exp}$, and
$z_{\rm exp}$
to approach $+\half$, $0$, and $+\half$, respectively.

Now suppose that the spin is prepared by a technician who puts it in either a pure state with $\sigma_z=+1$, or a pure state with $\sigma_x=+1$, and 
{\it makes the same choice every time}.  We, however, are not aware 
of what her choice is.

If forced to assign a particular density matrix, we would have to choose
\eq{rhoodd}.  However, our situation is clearly different from what it was
in the previous example.  In the present case, repeated experiments
would {\it not\/} verify \eq{rhoodd}, but would instead converge on
either $x_{\rm exp}$ = 0 and $z_{\rm exp}=+1$, or
$x_{\rm exp}$ = +1 and $z_{\rm exp}=0$.
Therefore, in this case, it is more appropriate to assign a prior probability of one-half to 
$\rho=\half(1+\sigma_z)$ and a prior probability one-half to $\rho=\half(1+\sigma_x)$.   
Then, as data comes in,
we can update these probability assignments with Bayes' theorem,
as described in section \ref{qm}.

Thus, it is better to choose $p(\rho)d\rho$ when it is possible that there is 
something about the preparation procedure that consistently prefers a particular
direction in Hlibert space, but we do not know what that direction is.
Since this possibility can rarely be ruled out {\it a priori},
we are typically better served by choosing a prior probability $p(\rho)d\rho$, 
rather than a particular vaue of $\rho$ itself.

\section{Noninformative Priors for Density Matrices}
\label{nonin}

Suppose we have decided to choose a prior probability 
$p(\rho)d\rho$ for the density matrix $\rho$ of some quantum system.
How should we choose this probability?  

In the case where we have little or no
information about the quantum system, we would like to formulate the appropriate analog of the principle of indifference.
Consider a qu$n$it, a quantum system
whose Hilbert space has dimension $n$ that is known to us.  
(We will not consider the even more general problem where $n$
is unknown.)   We can always write the density matrix (whatever
it is) in the form 
\begin{equation}
\rho = U^{-1}\tilde\rho \,U,
\label{UrhoU}
\end{equation}
where $U$ is unitary with determinant one,
and $\tilde\rho$ is diagonal with nonnegative entries
$p_1,\ldots, p_n$ that sum to one.    There is a natural
measure for special unitary matrices, the Haar measure;
it is invariant under $U\to CU$, where $C$ is a constant special unitary matrix.
In the simplest case of $n=2$, we can parameterize $U$ as 
$U=e^{i\alpha_1\sigma_3} e^{i\alpha_2\sigma_2} e^{i\alpha_3\sigma_3}$,
with $0\le\alpha_1\le\pi$, $0\le\alpha_2\le\pi/2$, $0\le\alpha_3\le \pi$;
then the normalized Haar measure is 
$dU=\pi^{-2}\sin(2\alpha_2)d\alpha_1 d\alpha_2 d\alpha_3$.
This construction is extended to all $n$ in \cite{sudar}.

Suppose we know that the state of the quantum system is pure.
Then we can set $\tilde\rho_{ij}=\delta_{i1}\delta_{j1}$, and parameterize $\rho$
via $U$.  Then it is natural to choose $d\rho=dU$ and $p(\rho)=1$, because
this is the only choice that is invariant under unitary rotations in Hilbert space.


Now consider the more general case where we do not have information about 
the purity of the system's quantum state.  Following \cite{meas}, 
we define the volume element via
\begin{equation}
d\rho \equiv dU\kern0.5pt dF,
\label{drho}
\end{equation}
where $dU$ is the normalized Haar measure for $U$, and
\begin{equation}
dF = (n{-}1)!  \,\delta(p_1+\ldots+p_n-1)dp_1\ldots dp_n
\label{dF}
\end{equation}
is a normalized measure for the $p_i$'s that we will call
the {\it Feynman measure\/} (because it appears in
the evaluation of one-loop Feynman diagrams).   
\Eq{dF} assumes that each $p_i$
runs from zero to one; then \eq{UrhoU} is an overcomplete construction,
because $U$ can rearrange the $p_i$'s.  This is easily fixed by imposing
$p_1\ge\ldots\ge p_n$, and multiplying $dF$ by $n!$.
However, \eq{dF} as it stands is easier to write and think about;
the overcompleteness of this construction of $\rho$ causes no harm.

In the case $n=2$, we previously chose $d\rho=dV=(3/4\pi)dx\,dy\,dz$ for
the parameterization of \eq{rhoxyz}.  In this case, the eigenvalues of $\rho$
are $\half(1+r)$ and $\half(1-r)$, with $0\le r\le 1$.  
After integrating over $U$, $dV\to 3 r^2\,dr$;
in comparison, $dF=dr$ for this case.  

The purity of a density matrix $\rho$ can be paramertized by ${\rm Tr}\,\rho^2$,
which for $n=2$ is $\half(1+r^2)$.  Thus the volume measure $dV$
is more biased towards pure states than is the Feynman measure $dF$;
we have $dV=3(2\,{\rm Tr}\,\rho^2-1)dF$.  

In general, we can accomodate any such bias by taking $p(\rho)d\rho$ to be 
of the form
\begin{equation}
p(\rho)d\rho = p({\rm Tr}\,\rho^2)dU\kern0.5pt dF ,
\label{prho}
\end{equation}
where $p(x)$ is an increasing function if we are biased towards having a pure state.
For $n>2$, we can take $p$ to be a function of ${\rm Tr}\,\rho^k$ for $2\le k\le n$.
Arguments in favor of various choices of $p$ have been 
put forth \cite{rhop}, but no single choice seems particularly compelling.
Of course, once we have done enough experiments,
our original biases become largely irrelevant, as we saw in section \ref{qm}.

\section{Conclusions}
\label{con}

We have argued that, in a Bayesian framework, the nature of our
ignorance about a quantum system can often be more faithfully 
represented by a prior probability $p(\rho)d\rho$
over the range of allowed density matrices, 
rather than by a specific choice of density matrix.
This method is particularly appropriate when (1) the preparation procedure
may favor a direction in Hilbert space, but we do not know what
that direction is, and (2) we can recreate the preparation procedure 
repeatedly, and perform measurements of our choice on each prepared
system.  In this case, as data comes in, we use Bayes' theorem to update $p(\rho)d\rho$.  Eventually, all but strongly biased observers (who can be identified {\it a priori\/} by an examination of their choice  of prior probability) will be convinced of the values of the quantum probabilities.
In this way, initially subjective probability assignments become more and
more objective.

In choosing $p(\rho)d\rho$, we can use the principle of indifference, 
applied to the unitary symmetry of Hilbert space, to reduce the problem to one
of choosing a probability distribution for the eigenvalues of $\rho$.  There is, however, no compelling rationale for any particular choice; in particular, we must decide how biased we are towards pure states.

\begin{acknowledgments}

I am grateful to Jim Hartle for illuminating discussions, and prescient comments on earlier drafts of this paper.  This work was supported in part by NSF Grant No.~PHY00-98395.

\end{acknowledgments}


\begin{thebibliography}{99}

\bibitem{cfs}
C.~M. Caves, C.~A. Fuchs, and R.~Schack, 
``Quantum probabilities as Bayesian probabilities,''
Phys.~Rev.~A {\bf 65}, 022305 (2002) 
[quant-ph/0106133].

\bibitem{definpp}
B.~de~Finetti, 
``Probabilities of probabilities: a real problem or a misunderstanding?''
in \textit{New Developments in the Application of Bayesian Methods},
A.~Aykac and C.~Brumat, eds. (North Holland, 1977).

\bibitem{dec}
R.~W.~Goldsmith and N.-E.~Sahlin, 
``The role of second-order probabilities in decision making,''
in \textit{Analysing and Aiding Decision Processes},
P.~Humphreys, O.~Svenson, and A.~Vari, eds.
(North Holland, 1983).

\bibitem{defin}
B.~de~Finetti, \textit{Theory of Probability} (Wiley, 1990).

\bibitem{qdef}
R.~L. Hudson and G.~R. Moody, 
``Locally normal symmetric states and an analogue of de Finetti's theorem,'' 
Z. Wahrschein.\ verw.\ Geb.\ {\bf 33}, 343 (1976);
C.~M. Caves, C.~A. Fuchs, and R.~Schack, 
``Unknown quantum states:\ the quantum de Finetti representation,'' 
J. Math. Phys. {\bf 43}, 4537 (2002)
[quant-ph/0104088].

\bibitem{qbayes}
R.~Schack, T.~A.~Brun, and C.~M.~Caves,
``Quantum Bayes rule,''
Phys. Rev. A {\bf 64}, 014305 (2001)
[quant-ph/0008113].

\bibitem{sudar}
T.~Tilma and E.~C.~G.~Sudarshan,
``Generalized Euler angle parametrization for SU($N$),''
J. Phys. A: Math. Gen. {\bf 35}, 10467 (2002) 
[math-ph/0205016].

\bibitem{meas}
K.~Zyczkowski, P.~Horodecki, A.~Sanpera, and M.~Lewenstein, 
``Volume of the set of separable states,''
Phys. Rev. A {\bf 48}, 883 (1998) 
[quant-ph/9804024].

\bibitem{rhop}
M.~J.~W.~Hall,
``Random quantum correlations and density operator distributions,''
Phys. Lett. A {\bf 242}, 123 (1998)
[quant-ph/9802052]; 
P.~B.~Slater,
``Comparative noninformativities of quantum priors based on monotone metrics,''
Phys. Lett. A {\bf 247}, 1 (1998)
[quant-ph/9703012];
``Monotonicity properties of certain neasures over the two-level quantum systems,''
Lett. Math. Phys. {\bf 52}, 343 (2000)
[quant-ph/9904014].




\end{thebibliography}
\end{document}